%% file: paper.tex
\documentclass[journal=jpclett,manuscript=letter,layout=traditional]{achemso}

\usepackage{graphicx}% Include figure files
\usepackage{dcolumn}% Align table columns on decimal point
\usepackage{bm}% bold math
\usepackage{glossaries}

\include{cmd}

\author{Tobias Sch\"afer}
\email{tobias.schaefer@tuwien.ac.at}
\affiliation{Institute for Theoretical Physics, TU Wien, Wiedner Hauptstraße 8-10/136, A-1040 Vienna, Austria}

\title{Ground-States for Metals from Converged Coupled Cluster Calculations}

\begin{document}

\begin{tocentry}

\includegraphics[width=1.0\linewidth]{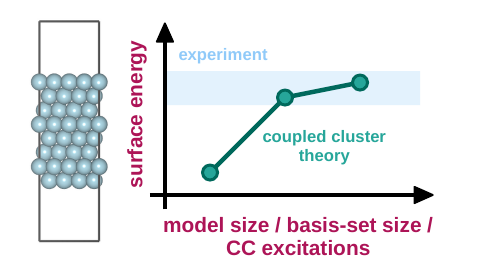}

\end{tocentry}

\begin{abstract}
Many-electron correlation methods offer a systematic approach to predicting material properties with high precision.
However, practically attaining accurate ground-state properties for bulk metals presents significant challenges.
In this work, we propose a novel scheme to reach the thermodynamic limit of the total ground-state energy of metals using coupled cluster theory.
We demonstrate that the coupling between long-range and short-range contributions to the correlation energy is sufficiently weak, enabling us to restrict long-range contributions to low-energy excitations in a controllable way.
Leveraging this insight, we calculate the surface energy of aluminum and platinum (111), providing numerical evidence that coupled cluster theory is well-suited for modeling metallic materials, particularly in surface science.
Notably, our results exhibit convergence with respect to finite-size effects, basis-set size, and coupled cluster expansion, yielding excellent agreement with experimental data.
This paves the way for more efficient coupled cluster calculations for large systems and a broader utilization of the theory in realistic metallic models of materials.
\end{abstract}

\maketitle

%\section{Introduction}

\textit{Introduction.---}
Accurately predicting quantum mechanical ground-state properties is fundamental to materials modeling and requires the most advanced computational methods. Systematically improvable many-electron correlation methods play a pivotal role, with the family of \gls{cc} methods standing out as a prominent and widely recognized approach~\cite{Coester1960,Cizek1966,Cizek1971,Zhang2019}. 
Within \gls{cc} methods, the true many-electron wavefunction $|\psi\rangle$ is approximated through excitations of a mean-field Slater determinant $|\phi\rangle$---mostly from \gls{hf} theory---with accuracy systematically enhanced through consideration of higher-order excitations $T=T_1+T_2+\dots$ by means of an exponential ansatz $|\psi\rangle = \EuE^T |\phi\rangle$.

In materials science, the accuracy is often considered in the notion of a hypothetical three-dimensional parameter space encompassing the employed simulation cell size, the one-electron basis-set size, and the order of excitation operators in the \gls{cc} method. 
Given a desired accuracy necessitates convergence of all three parameter axes, encompassing the \gls{tdl}, the \gls{cbs}, and the limit of excitation orders, respectively.
Due to the steep scaling of the computational cost along all three axes, achieving converged coupled cluster calculations is often practically infeasible for extended systems like solids~\cite{Booth2012,Zhang2019}. 
For metals, in particular, the goal of materials modeling from converged \gls{cc} results often remains out of reach~\cite{Mihm2021,Neufeld2022}.
The limited number of fully converged \gls{cc} results leaves gaps in our understanding of the theory's performance for materials.

In this work, we present a novel finite-size correction scheme to reach the \gls{tdl} of CC ground-state energies for real metals. 
This scheme, combined with recent methodological advancements, forms a robust and massively parallelized computational framework for calculating converged \gls{cc} ground-state properties across all three hypothetical parameters.
We apply this framework to determine the surface energy of aluminum and platinum in the (111) termination.

Surface properties are crucial in various research fields, including heterogeneous catalysis, energy storage, or corrosion, as materials primarily interact with their environment at their surfaces.
While \gls{dft} is undeniably the leading method used in materials modeling, existing approximations of the exchange-correlation functional often fail to reliably predict surface energies~\cite{Miller2009}.
More advanced methods like the \gls{rpa} show improved performance over \gls{dft} for metal surface energies~\cite{Schimka2010,Patra2017,Schmidt2018}, but suffer from known issues, introducing uncertainties in the results.
Notably, \gls{rpa} suffers from unphysical self-correlation, which only cancels out for energy differences of similar electronic structures~\cite{Gruneis2009,Bates2012,Schafer2021c,Ruan2021}.
A study systematically approaching the exact solution of the many-electron Schr\"odinger equation for  surface models is still lacking.

\textit{Methods.---}
Our approach to systematically converge \gls{cc} theory combines our novel finite-size correction scheme to reach the \gls{tdl} with several recently published methodological advancements.
Starting from a plane-wave basis representation of the \gls{hf} orbitals from the \gls{vasp}~\cite{Kresse1993,Kresse1996a,Kresse1996b}, we efficiently reach the \gls{cbs} through a transformation to a more compact natural orbital basis~\cite{Gruneis2011a,Ramberger2019} in combination with a highly effective focal point correction scheme~\cite{Irmler2021}.
An efficient solution to the long-standing problem of accurately capturing three-electron correlation effects in metals without running into an infrared catastrophe~\cite{Shepherd2013,Neufeld2023}---as is the case in CCSD(T) using the perturbative triples treatment (T)~\cite{Raghavachari1989}---has recently been proposed by us~\cite{Masios2023}.
In this new approach, denoted as CCSD(cT), significant three-electron screening effects are accounted for, preventing the infrared catastrophe and providing an accurate estimate for coupled cluster calculations with single, double, and triple excitations~\cite{Schafer2024X}.
Using our massively parallelized cc4s code~\cite{cc4s}, we show that for the surface energy of aluminum and platinum (111) convergence of the excitation orders is reached using CCSD(cT).
Furthermore, two points are crucial for the successful calculation of ground-states of metals in this work.
First, we employ a Monte-Carlo integration technique to sample the Brillouin zone, which allows us to converge the \gls{hf} and \gls{cc} energy of a finite simulation cell without relying on high-symmetry points~\cite{Lin2001}.
This so-called twist-averaging technique is important as it prevents degenerate \gls{hf} energies and ensures purely integer orbital occupation numbers~\cite{Mihm2021}.
Additionally, we utilize a recently published sampling technique for the reciprocal Coulomb potential, which enables converged \gls{hf} and \gls{cc} ground-state energies in anisotropic simulation cells~\cite{Schafer2024}, as necessary for modeling surface slabs.
In the following, we introduce our novel finite-size correction scheme.

\textit{Achieving the \gls{tdl} of real metals.}---The correlation energy per unit, such as an atom or unit cell, of a periodic system can be expressed as
\begin{equation}
E_c = \int  v(\bm q) \, S(\bm q) \; \D^3 q \;, \label{eq:Ec}
\end{equation}
where $v(\bm q) = 4\pi / \bm q^2$ is the Fourier transform of the Coulomb potential $1/|\bm r|$  and we call $S(\bm q)$ the transition structure factor~\cite{Gruber2018} per unit. 
The transition structure factor can be interpreted as the Fourier transform of the transition pair correlation function~\cite{Irmler2019}, encompassing the physical information about the many-electron correlations in the system and is determined by the many-electron Schr\"odinger equation.
More formally, it can also be considered as the partial derivative of the correlation energy $E_c$ with respect to the reciprocal potential $v(\bm q)$. 
In this work $S(\bm q)$ serves as the main quantity to study different correlation lengths as long-range and short-range correlation effects are decoded in small and large magnitudes of the transition vectors $\bm q$, respectively~\cite{Liao2016}.
This is a consequence of the fact that we consider the Fourier transformations of real space quantities, i.e. $\bm q$ is the conjugate variable of $\bm r$.
In plane wave and real-space grid based approaches in particular, $\bm q$-dependent quantities are inherently accessible.
In any finite-cell simulation, the reciprocal transition vectors $\bm q$, also known as momentum transfer vectors, can be decomposed into the sum of a reciprocal lattice vector and a difference vector of two grid points sampling the first \gls{bz}~\cite{Liao2016}.
Hence, a finite grid of $\bm q$ vectors is defined by the sampling grid of the \gls{bz}, the unit cell, and a cutoff parameter which specifies the largest $|\bm q|$ to be considered.
In a practical computation the integral of Eq. (\ref{eq:Ec}) turns into a sum over this grid.
We note that the singularity at $\bm q = 0$, although integrable, requires careful handling~\cite{Schafer2024,Xing2024}.

\begin{figure}
    \centering
    \includegraphics[width=0.8\linewidth]{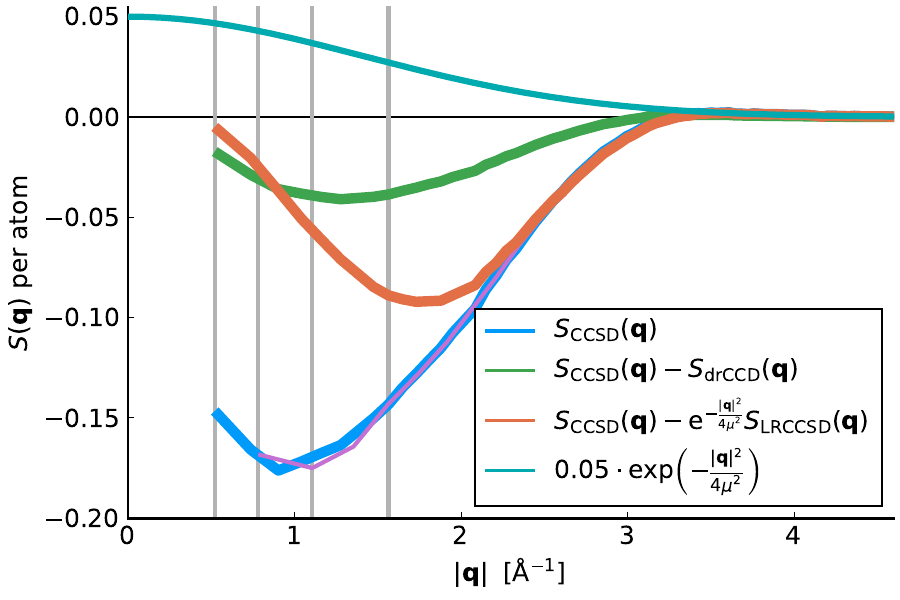}
    \caption{
    Transition structure factors and differences of structure factors in reciprocal space of metallic aluminum using finite supercells of 108 atoms.
    The thin purple line shows $S_\text{CCSD}(\bm q)$ for a supercell of 32 atoms.
    The structure factors are spherically averaged, hence the area under each curve provides an estimate of the corresponding correlation energy.
    The four gray vertical lines indicate the smallest $|\bm q|$ vector for supercells with 108, 32, 16, 4 atoms (left to right). Also the exponential damping function for the long-range potential is shown.}
    %The fact that CCSD-LRCCSD might be due to the limited number of stochastic twists, leading to remaining noise in the SF result. Making the cell larger, would include smaller q-points, however, there the difference CCSD-LRCCSD is almost zero, so almost no correlation energy will be added. Hence the correction is almost aturated.}
    \label{fig:SF}
\end{figure}

In coupled cluster theory the structure factor $S(\bm q)$ is accessible through the double excitation amplitudes $t^{ab}_{ij}$ as
\begin{equation}
S(\bm q) =  \sum_{ijab} t^{ab}_{ij} \big( 2C^a_i(\bm q) C^j_b(\bm q)^* -  C^b_i(\bm q) C^j_a(\bm q)^*\big)\;,
\end{equation}
where $C^a_i(\bm q)$ are the Fourier transforms of the overlap densities $\varphi_i^*(\bm r)\varphi_a(\bm r)$ of the occupied and unoccupied (virtual) mean-field orbitals $i$ and $a$, respectively.
The double excitation amplitudes $t^{ab}_{ij}$ are solutions of the \gls{cc} amplitude equations which are derived from the \gls{cc} ansatz mentioned in the introduction and the time-independent Schr\"odinger equation.
Considering \gls{ccsd} the amplitude equations are basically self-consistent equations, 
consisting of contractions of the Coulomb tensor $v^{ab}_{ij}$, the amplitudes itself, as well as \gls{hf} orbital energy differences $\Delta^{ij}_{ab} = \varepsilon_i + \varepsilon_j - \varepsilon_a - \varepsilon_b$,
\begin{align}
t^{ab}_{ij} =  \big( &v^{ab}_{ij} + v^{ak}_{ic} t^{cb}_{kj} + v^{kb}_{cj} t^{ac}_{ik} +  v^{ab}_{cd} t^{cd}_{ij} + v^{kl}_{cd} t^{ac}_{ik} t^{db}_{lj} - v^{kb}_{id} t^{ad}_{kj} + ... \big) / \Delta^{ij}_{ab} \;,
\label{eq:ccsdamp}
\end{align}
Summation  over repeated indices is assumed.
We only show selected terms of the \gls{ccsd} equations (for the full set of terms we refer to Ref. \citenum{Shavitt2010}) to illustrate that $S(\bm q)$ explicitly depends on both the $\bm q$ grid and the amplitudes, while the amplitudes implicitly depend on the $\bm q$ grid through the Coulomb integrals.
Under periodic boundary conditions, the Coulomb integrals $v^{ab}_{ij}$ can be calculated as a sum over the grid,
\begin{equation}
v^{ab}_{ij} =  \sum_{\bm q}   v(\bm q) C^i_a(\bm q) C^b_j(\bm q)^* \;,
\end{equation}
a formulation naturally accessible in a plane wave basis. 
In practice, the sum over $\bm q$ includes a weighting factor that depends on the chosen integration grid, but this is omitted here for brevity.

In this context, the finite-size error in the correlation energy---i.e. deviations of the correlation energy from its value in the thermodynamic limit---can be attributed to the grid's coarseness leading to missing information near $\bm q = 0$.
This finite-size error diminishes as the sampling of the \gls{bz} becomes infinitely dense, or equivalently, as the simulation cell becomes infinitely large. 
%It is important to note that as $\bm q \rightarrow 0$, we have $v(\bm q)S(\bm q) = 0$, reflecting the fact that no correlation energy is contributed by infinite correlation lengths.

Few strategies exist for correcting finite-size errors~\cite{Nusspickel2022,Ye2024,Xing2024}, with cell-size extrapolation techniques~\cite{Mihm2023} being among the most commonly used.
Alternatively, the structure factor offers another method to estimate these corrections.
Liao et al. \cite{Liao2016} introduced a structure factor interpolation scheme. 
This method approximates the limit of an infinitely dense $\bm q$ grid (the \gls{tdl}) using a tricubic interpolation technique based on $S(\bm q)$ on the finite grid.

%Here, we aim to calculate the missing q contribution using a multiscale approach in which we 
Inspired by the structure factor based approach, we employ a multiscale approach here, which aims at approximating the small $\bm q$ contributions missing in the grid. 
To this end, we probe the coupling strength of small and large $\bm q$ contributions.
In other words, can $S(\bm q)$ be accurately calculated for small $\bm q$ while neglecting larger $\bm q$?
One way to scrutinize this is by solving the \gls{ccsd} equations using a potential which smoothly damps large transition momentum vectors, like $4\pi\exp(-\frac{\bm q^2}{4\mu^2})/\bm q^2$.
In real space, this corresponds to a long-range (LR) potential using the error function $1/|\bm r| \rightarrow \text{erf}(\mu |\bm r|)/|\bm r|$.
We use the prefix LR to indicate solutions of the CC equations employing a long-range potential, such as LRCCSD for long-range CCSD.
For simplicity, we chose a fixed parameter setting of $\mu = 1.0 \text\AA^{-1}$ for the error function throughout this work.

Fig. \ref{fig:SF} shows the structure factor $S(\bm q)$ at the level of \gls{ccsd} for bulk aluminum in the fcc structure. 
For the calculation a $3\times3\times3$ supercell of the conventional unit cell was employed, resulting in a finite-size model containing 108 atoms, as visualized in the \gls{si}~\cite{si}.
The smallest finite transition momentum vector is determined by this choice. 
The difference between structure factors of the CCSD and the LRCCSD calculation approaches zero for small $|\bm q|$.
Hence, both structure factors describe similar long-range effects in the correlation energy calculated via Eq. (\ref{eq:Ec}).
This indicates a weak coupling between long-range and short-range correlation effects, supporting a multiscale approach that aims for the \gls{tdl} while neglecting the short-range.

\begin{figure}
    \centering
    \includegraphics[width=0.8\linewidth]{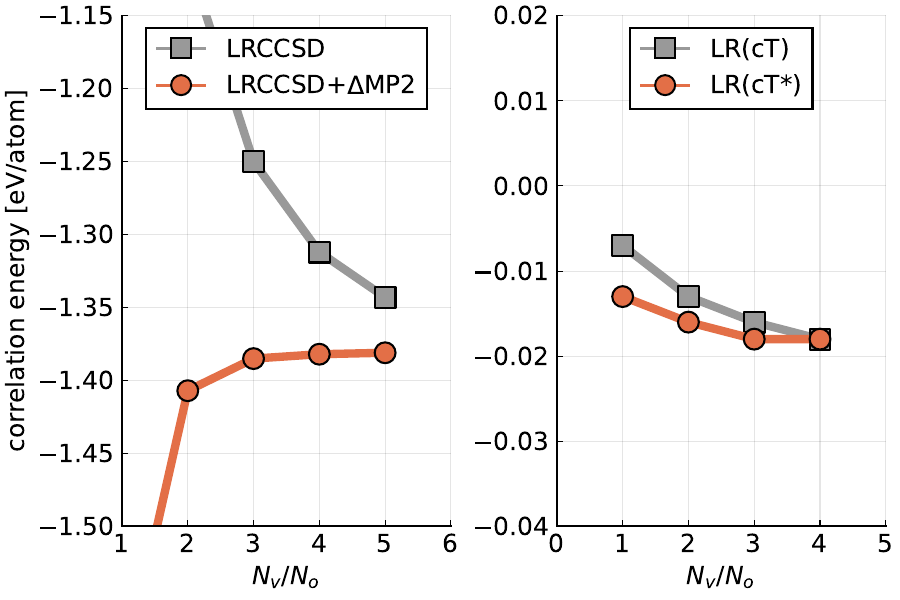}
    \caption{Rapid convergence of the total LRCCSD (left) and LR(cT) (right) correlation energy of metallic aluminum with respect to the number of basis funcitions per occupied orbitals, $N_v/N_o$. The basis functions are given by approximate natural orbitals~\cite{Gruneis2011}.
    An even faster convergence can be achieved by adding a correction based on the complete basis-set limit of the MP2 level. Details on $\Delta$MP2 and LR(cT*) are provided in the \gls{si}~\cite{si}.
    In this work we chose $N_v/N_o=3$.} 
    \label{fig:bsc}
\end{figure}

Additionally, the long-range potential offers a significant computational advantage by enabling a substantial reduction in the basis-set size, thereby greatly decreasing the computational workload.
Consistent with previous observations, a much smaller basis-set can be used to reach the \gls{cbs}~\cite{Toulouse2004,Bruneval2012,Giner2018,Riemelmoser2020}.
Figure \ref{fig:bsc} shows the convergence rate for the correlation energy using the long-range potential. 
This is because the electron-electron cusp is less sharp for a long-range potential compared to a Coulomb potential~\cite{Franck2015}.
Thus, we can systematically control the space of low-energy excitations required to accurately capture the long-range contributions to the ground-state energy.
Note that the convergence rate to the \gls{cbs} depends on the choice of $\mu$, converging faster/slower for smaller/larger $\mu$.
In our setting, only about 3 unoccupied orbitals per occupied orbital are necessary to achieve an accuracy well below $5 \,\text{meV}$ per atom for the total energy.
Such an accuracy is impossible to attain with current computational resources for total correlation energies using the Coulomb potential.
Typically, around 20 or more optimized Gaussian basis functions per occupied orbital---in combination with effective correction schemes like the explicitly correlated F12 method~\cite{Grueneis2017}---are required , even for energy differences that benefit from significant error cancellation, to reach chemical accuracy (1 kcal/mol or about 43 meV) for observables like atomization or reaction energies~\cite{Irmler2021}.

To estimate the reduction in computational cost, note that reducing the basis-set size by a factor of $x$ decreases the memory footprint by a factor of $x^2$ and the computation time by a factor of $x^4$ at the \gls{ccsd} level. For higher orders of \gls{cc} theory, the savings are even more pronounced.
The computation time in dependence of the cell size is provided in the \gls{si}~\cite{si}.

\begin{figure}
    \centering
    \includegraphics[width=0.8\linewidth]{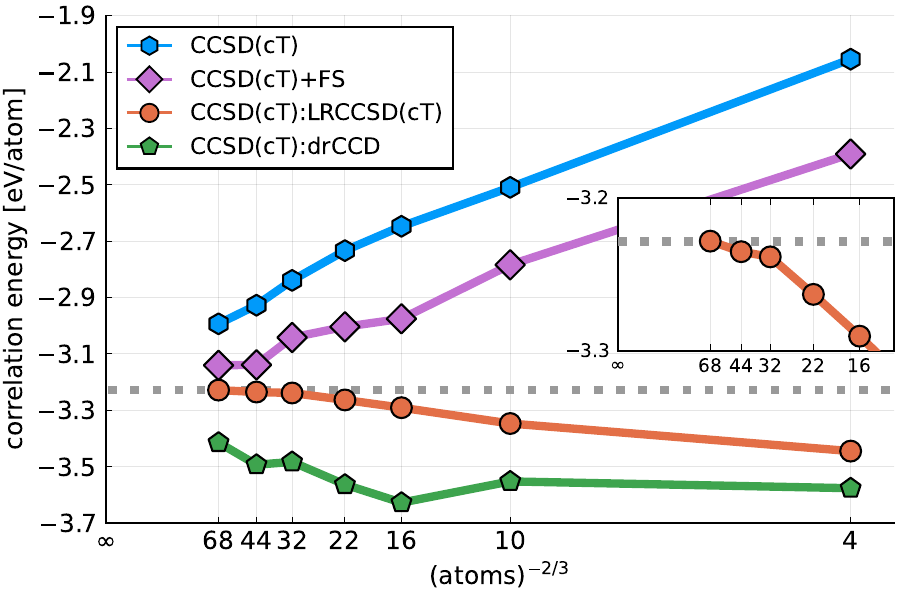}
    \caption{Approaching the total correlation energy per atom of metallic aluminum at the level of \gls{ccsdpct} using different finite-size corrections schemes. The correction schemes are introduced in the main text. The finite-size of the periodic model is indicated by the number of atoms on the horizontal axes. The dashed line shows the extrapolated CCSD(cT) energy.}
    \label{fig:AlTDL}
\end{figure}

Based on these observations, we introduce a finite-size correction scheme to approximate the \gls{tdl} for real metals modeled by finite supercells. 
The main idea is to shift the task of reaching the \gls{tdl} of the CC ground-state energy to the less computationally demanding long-range contribution using the following equation,
\begin{align}
E^\text{TDL}_\text{CC} \approx  
E^\text{TDL}_\text{CC:LRCC} &=
E^\text{finite}_\text{CC}
-E^\text{finite}_\text{LRCC}
+ E^\text{TDL}_\text{LRCC} \nonumber\\
&= \sum_{\bm q} v(\bm q) \big( S^\text{finite}_\text{CC}(\bm q) - \EuE^{-\frac{\bm q^2}{4\mu^2}} S^\text{finite}_\text{LRCC}(\bm q) \big) + E^\text{TDL}_\text{LRCC}  \, .
\label{eq:fsc}
\end{align}
The \gls{tdl} in $E^\text{TDL}_\text{LRCC}$ can be estimated using existing techniques, such as cell size extrapolations or structure factor interpolations, but with greatly reduced computational cost.
Computational details used for the following results can be found in the \gls{si}~\cite{si}.

Figure \ref{fig:AlTDL} illustrates the performance of this finite-size correction for the total CCSD(cT) energy of metallic aluminum, denoted as CCSD(cT):LRCCSD(cT).
The energy difference between CCSD(cT) and LRCCSD(cT) saturates quickly with increasing cell size, making this correction scheme more effective than the one based on the structure factor interpolation~\cite{Liao2016}, denoted as CCSD(cT)+FS.
It must be fairly noted that the used implementation of the FS correction is not fully warranted for metallic systems, as it assumes a quadratic behaviour of the structure factor around $\bm q = 0$.
This is the correct bevavior for insulating systems, leading to a finit-size error scaling of $N^{-1}$.
Nevertheless, it has been shown that metals require a linear interpolation of the structure factor around small transition vectors, implying a $N^{-2/3}$ behaviour~\cite{Mihm2023}, where $N$ is the number of atoms in the supercell.

Since the (cT) contribution is evaluated in a single-shot calculation, the finite-size effects can be corrected separately in the sense of CCSD(cT):LRCCSD(cT) = CCSD:LRCCSD + (cT):LR(cT).
The separate convergence is documented in the \gls{si}~\cite{si} and shows that both the total correlation energy and the finite-size error are dominated by the CCSD contribution. 
While the CCSD contribution to the correlation energy is of the order of $-3\,\text{eV}$, the (cT) contribution is of the order of $-0.1\,\text{eV}$.
The \gls{si} also shows that the long-range based correction outperforms the extrapolation technique at the level of \gls{ccsd}.

Another advantage of the proposed correction scheme based on the long-range potential is that it accounts for the so called minimum drifting of the structure factor, which was already described by Weiler \emph{et. al}~\cite{Weiler2022}.
The drifting of the characteristic minimum of the structure factor can be observed in Fig. \ref{fig:SF}, when comparing the CCSD structure factor of the 108 atoms cell (blue) with 32 atoms cell (thin purple line).
The minimum shifts to the left for increasing cell sizes (i.e. finer $\bm q$ grids).
Weiler \emph{et al.} suggest that the drifting is due to relaxations of the CC amplitudes from Eq. (\ref{eq:ccsdamp}) as the grid is refined.
The CC:LRCC method avoids this issue, as the effects from the drifting can be assumed to cancel out in the difference $S^\text{finite}_\text{CC}(\bm q) - \EuE^{-\frac{\bm q^2}{4\mu^2}}S^\text{finite}_\text{LRCC}(\bm q)$.

Additionally, a correction based on the \gls{drccd} theory~\cite{Scuseria2008,Gruneis2009}, denoted as CCSD(cT):drCCD is considered.
It is equivalently defined via Eq. (\ref{eq:fsc}) by replacing LRCC with drCCD.
This correction draws from our previous works, where we showed that the long-range contributions of fragment-based applications of CCSD(T) theory can be effectively approximated with ring-like terms of \gls{ccsd} for materials with a gap~\cite{Schafer2021a,Schafer2021b}.
However, as apparent from Fig. \ref{fig:SF} and \ref{fig:AlTDL} this does not seem to apply for the total correlation energy of metals.
The difference of the structure factors of CCSD and drCCD exhibits a slower decay to zero for decreasing $|\bm q|$ (increasing cell sizes) as compared to the difference between CCSD and LRCCSD.

\begin{figure}
    \centering
    \includegraphics[width=0.8\linewidth]{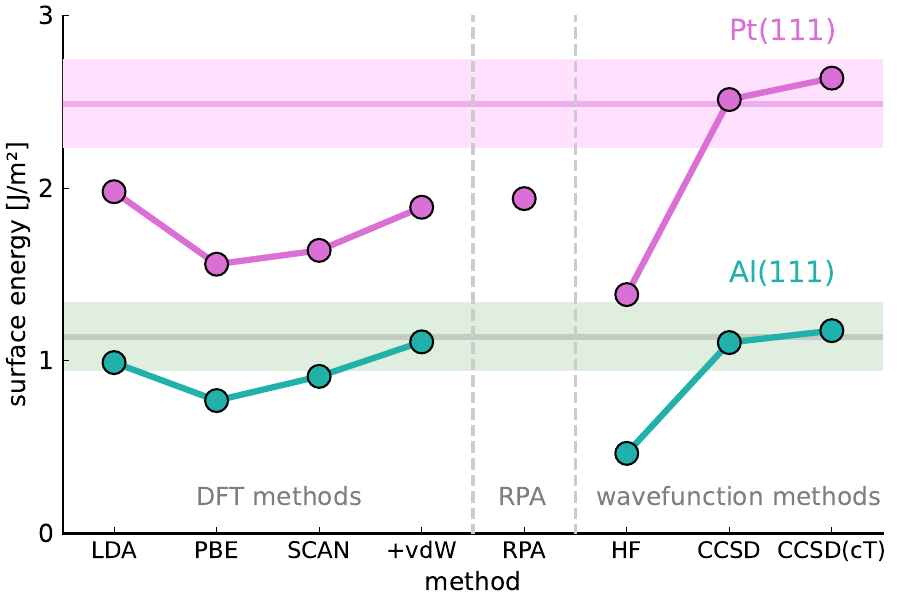}
    \caption{Surface energy of aluminum and platinum in the (111) termination from various methods. The experimental results (horizontal lines) and their uncertainties (lighter area), as well as the values of the different DFT functionals are taken from Ref. \cite{Patra2017}. 
    Here +vdW denotes SCAN+vdW as described in the text.
    The \gls{rpa} result is taken from Ref.~\cite{Schimka2010}}
    \label{fig:gamma}
\end{figure}

\textit{Surface energy.}---
We now turn to the surface energy of aluminum and platinum.
The surface of a material determines its properties for scientific and industrial applications.
The presented finite-size correction scheme enables us to study the convergence of systematically improved wavefunction methods by increasing the \gls{cc} excitations orders---HF, CCSD, CCSD(cT)---for surface energies in the \gls{tdl}.
The results are show in Fig. \ref{fig:gamma}.
While well-established \gls{dft} functionals, such as LDA~\cite{Perdew1992}, PBE~\cite{Perdew1996}, SCAN~\cite{Sun2015}, as well as SCAN corrected by van-der Waals (vdW) contributions (here rVV10, i.e. revised Vydrov–Van Voorhis 2010~\cite{Peng2016}) are at least close to the experimental values for Al(111), they severely underestimate the surface energy of Pt(111).
In both cases, the systematically improved wavefunction methods show smooth convergence, with only minor corrections arising from contributions beyond \gls{ccsd}.
This suggests that surface energies converge rapidly with respect to the excitation order in \gls{cc} theory.
Notably, this outcome was achieved by employing a well-defined method to capture triple excitation effects, here (cT), which does not suffer from an infrared catastrophe as the perturbative triples approach, (T), does.
%Depending on the desired accuracy, this convergence indicates that \gls{ccsd} alone may be sufficient.
For both materials, \gls{ccsdpct} demonstrates excellent agreement with experimental values. 
In units of $\text{J/m}^2$ we find surface energies of $1.17$ for aluminum and $2.65$ for platinum using \gls{ccsdpct}, compared to experimental values~\cite{Patra2017} of $1.14\pm 0.20$ and $2.49\pm 0.26$, respectively.
The residual numerical uncertainty in the computed CCSD(cT) estimates is determined in the \gls{si} to be less than $0.1\,\text{J/m}^2$.
It is important to note that the available experimental results are relatively old, extrapolated to $T=0\,\text{K}$ from surface tension measurements of the liquid phase, and specific terminations like (111) are not directly accessible.
Despite these limitations, the experimental results provide valuable, albeit uncertain, estimates for the low-energy faces of bulk crystals which are frequently referenced in notable studies~\cite{Vitos1998,Miller2009,Schimka2010,Patra2017,Schmidt2018}.
%We note, that the experimental surface energies are \cite{Patra2017}
Additional deviations from experimental values may arise from effects not considered in this work, including contributions from frozen core electrons, relativistic effects of the valence electrons, and ionic relaxation of the slabs. 
Previous \gls{dft} studies, however, have shown that ionic relaxation effects are very small~\cite{Miller2009}.

Notably, the finite-size correction for the \gls{cc} correlation contribution to the surface energy in the $xy$ direction (parallel to the surface) is virtually zero for the $2\times 2$ slab of Al(111).
This is dicussed in the \gls{si}~\cite{si}.
At first glance, this might seem contradictory to reports that show slower convergence in the $xy$ direction~\cite{Sheldon2021}.
However, the separated treatment of electrostatic \gls{hf} and correlation contributions to the surface energy reveals that the correlation contribution converges quickly with respect to the $xy$ dimension of the slab.
Twist-averaging plays an important role in this, as it smoothes the convergence compared to the more erratic convergence observed with $\bm\Gamma$-centered meshes for the \gls{bz} sampling.
Thus, long-range effects in the $z$ direction (normal to the surface) dominate the correlation part of the surface energy.
This outcome, revealed through our novel finite-size correction scheme, underscores its ability and would have been difficult to detect otherwise.
Details can be found in the \gls{si}~\cite{si}.

Our investigation also highlights a limitation of the finite-size correction based on the structure factor interpolation, above denoted as CCSD(cT)+FS. 
This scheme predicts a CCSD(cT) result of $0.5\,\text{J/m}^2$ for the surface energy of aluminum and $2.4\,\text{J/m}^2$ for platinum, considering $2\times 2$ surface slabs. 
The error introduced by this correction for aluminum arises from the interpolation of the structure factor in the $xy$ direction. 
As illustrated in the \gls{si}~\cite{si}, the smallest $\bm q$ vectors in the $xy$ direction do not reach the characteristic minimum of the structure factor, leading to inaccurate finite-size corrections of the +FS method when interpolating down to $\bm q = 0$.

\textit{Conclusion \& Outlook.}---
We introduce a novel finite-size correction scheme to enable coupled cluster theory for highly accurate materials modeling of metals.
Using the example of metal surface energies, which are highly relevant due to their wide range of applications, we demonstrated that this observable can be reliably reproduced with high precision for the first time.

Furthermore, we employed a recently published methodology for treating approximate triples correlation effects, denoted as CCSD(cT).
Our results add further evidence that this approximation is both accurate and practical for applications in metallic solids.

The proposed workflow to systematically converge \gls{cc} calculations for metals can be considerably accelerated even further, when combined with structure factor interpolation techniques and \textit{the shortcut to the \gls{tdl}} proposed in Ref. \cite{Mihm2021}.
Additionally, the weak coupling of different length scales in $\bm q$ space can serve as a foundation for designing novel reduced-cost algorithms.

This breakthrough paves the way for more efficient and more confident utilization of coupled cluster theory in materials science including metals, necessary for research areas such as the rational design of heterogeneous catalysts, the development of new functional materials, and the provision of highly-accurate benchmark results for machine learning techniques.

\bigskip 

This work was supported and funded by the Austrian Science Fund (FWF) [DOI: 10.55776/ESP335].
The computational results presented have been largely achieved using the Vienna Scientific Cluster (VSC-5).
Discussions with Evgeny Moerman, Johanna P. Carbone, Andreas Gr\"uneis, Alejandro Gallo, and Andreas Irmler are gratefully acknowledged.

\bibliography{refs}% Produces the bibliography via BibTeX.

\end{document}

%\title{Attaining chemically accurate ground-state properties of metals from converged coupled cluster calculations}
\title{Supplementary information for:\\Ground-States for Metals from Converged Coupled Cluster Calculations}

%\thanks{A footnote to the article title}%
\author{Tobias Sch\"afer}
\email{tobias.schaefer@tuwien.ac.at}
\affiliation{Institute for Theoretical Physics, TU Wien, Wiedner Hauptstraße 8-10/136, A-1040 Vienna, Austria}

\maketitle

\section{Computational details and workflow}

All \gls{hf} calculations are performed with the \gls{vasp}~\cite{Kresse1993,Kresse1996a,Kresse1996b} using a plane-wave cutoff of \texttt{ENCUT}=300 eV. The following pseudopotentials (\texttt{POTCAR} files) using the \gls{paw} method~\cite{Blochl1994} are employed: \texttt{PAW\_PBE Al\_GW} for aluminum providing 3 valence electrons and \texttt{PAW\_PBE Pt\_GW} for platinum providing 10 valence electrons.
The \gls{cc} calculations are performed with the CC4S code based on the interface from \gls{vasp}~\cite{cc4s}.
For the surface slabs, a recently published sampling technique for the Coulomb potential was used in order to achieve converged \gls{hf} and \gls{cc} iterations in strongly anisotropic simulation cells~\cite{Schafer2024}.
This sampling technique lets us use a $10\,\text\AA$ vacuum for the slab models, effectively eliminating interactions with periodic images in the $z$ direction due to the distance.

\subsection{Workflow}

\begin{figure}
    \centering
    \includegraphics[width=0.6\linewidth]{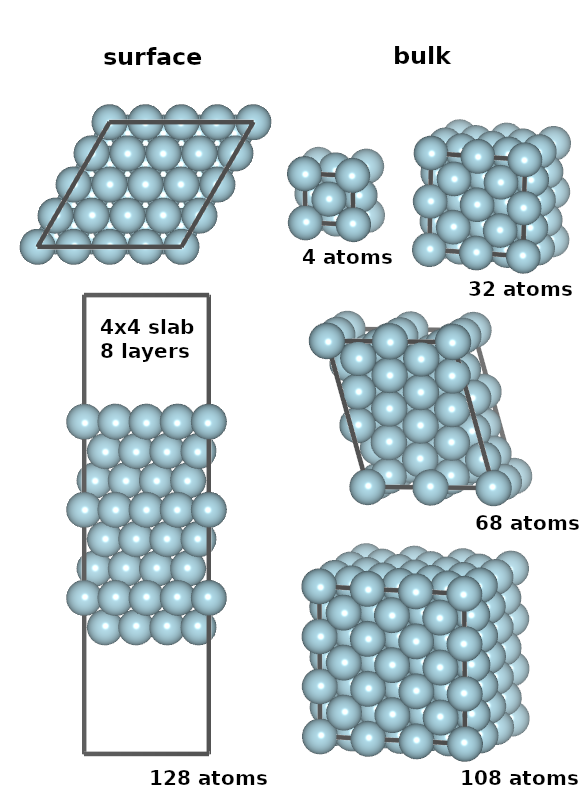}
    \caption{Selected models used in this work. Surface models are shown on the left, bulk models on the right. A vacuum of $10\,\text\AA$ is chosen for the surface slabs. A lattice constant of $4.018\,\text\AA$ is considered . The lattice vectors are shown as dark gray lines. Aluminum atoms are represented as space-filling blue balls.}
    \label{fig:models}
\end{figure}

In the following, we describe the general workflow for \gls{cc} calculations, performed for the main manuscript.
(VASP)~\cite{Kresse1993,Kresse1996a,Kresse1996b}  and the \texttt{Cc4s}~\cite{Gruber2018} code.
All correlation calculations are performed based on a single k-point sampling of the \gls{bz} using increasingly large supercells.
A selection of the used models for aluminum is visualized in Fig. \ref{fig:models}.
The following procedure is repeated for randomly chosen k-points, a Monte-Carlo technique called twist-averaging~\cite{Lin2001,Mihm2023}.

\begin{enumerate}

\item 
The \gls{hf} ground state is calculated. Both the occupied as well as all unoccupied orbitals and orbital energies are stored.

\item 
Approximate natural orbitals at the MP2 level are computed as described in Ref.~\cite{Gruneis2011}.
Natural orbitals are the eigenvectors of the one-electron reduced density matrix, with their corresponding eigenvalues referred to as occupation numbers.
Arranged by their occupation numbers, we truncated and recanonicalized the natural orbital basis by selecting a ratio $N_v/N_o$, where $N_o$ represents the number of occupied orbitals in the system and $N_v$ represents the number of chosen natural orbitals.
This basis of natural orbitals facilitates a much quicker convergence of the correlation energy with respect to $N_v$.

\item 
The MP2 energy is calculated in the CBS limit using the natural orbitals with $N_v/N_o = 100$.
This is necessary for basis set correction schemes to estimate the CBS limit of the CCSD and (cT) energies.
The basis set correction scheme (called focal point correction) is described in Ref. \cite{Irmler2021}.
For (cT) we use the (cT*) method in analogy to (T*)~\cite{Knizia2009} in order to estimate the \gls{cbs}.
The (cT*) estimate is obtained by rescaling (cT) with the relation between the small basis result of MP2 and the \gls{cbs} of MP2.

In the case of the long-range potential $N_v/N_o = 16$ is chosen. 

\item
For the coupled cluster calculations, a smaller basis, defined by $N_v/N_o = 8$, is selected. 
Basis set convergence is discussed below.
All Coulomb integrals, $V^{pq}_{sr}$, required by coupled cluster theory are calculated using the expression:
\begin{equation}
V^{pq}_{sr}=\sum_{F=1}^{N_F}{\Gamma^*}_s^{pF}{\Gamma}_{rF}^{q},
\end{equation}
where $p, q, r, s$ are indices for occupied or virtual orbitals.
$F$ denotes auxiliary basis functions obtained via singular value decomposition as described in Ref.~\cite{Hummel2017}.
Due to the significant vacuum in the surface slabs, the auxiliary basis set size can be considerably reduced without sacrificing the precision of the computed correlation energies.
The correlation energies are converged to within meV relative to the size of the optimized auxiliary basis set.

\item
The final coupled cluster calculations at various levels (drCCD, CCSD, and CCSD(cT)) are executed using the high-performance code cc4s.
%Up to 50 compute nodes with 128 cores each were utilized to implement our extensive computational parallelization approach.

\end{enumerate}

\subsection{Complete basis set limit of the long-range coupled cluster results}

As shown in the main manuscript the rapid convergence of the long-range coupled cluster energies with respect to the basis-set size $N_v/N_o$ can be accelerated by adding a correction based on the MP2 energy via
\begin{equation}
\text{LRCC}+\Delta\text{MP2} = \text{LRCC(small basis)} -  \text{LRMP2(small basis)} +  \text{LRMP2(CBS)}\;.
\end{equation}
We also define 
\begin{equation}
\text{LR(cT*)} = \text{LR(cT)(small basis)} \cdot \frac{\text{LRMP2(CBS)}}{\text{LRMP2(small basis)}}    \;. \label{eq:star}
\end{equation}
In this work we used a fix setting of $N_v/N_o=3$ for CC and $N_v/N_o=16$ for MP2.

\section{Calculation of the surface energy}

\begin{figure}
    \centering
    \includegraphics[width=0.49\linewidth]{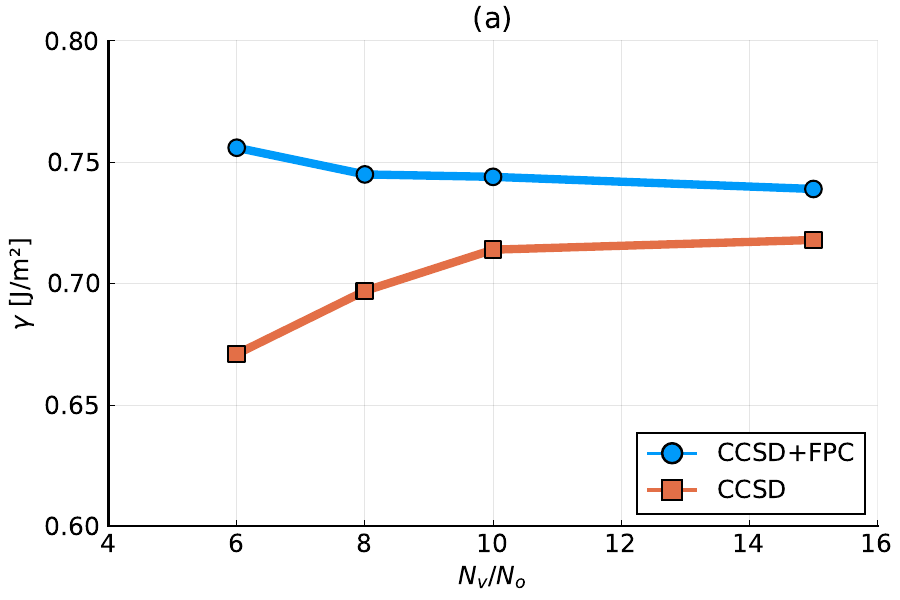}
    \includegraphics[width=0.49\linewidth]{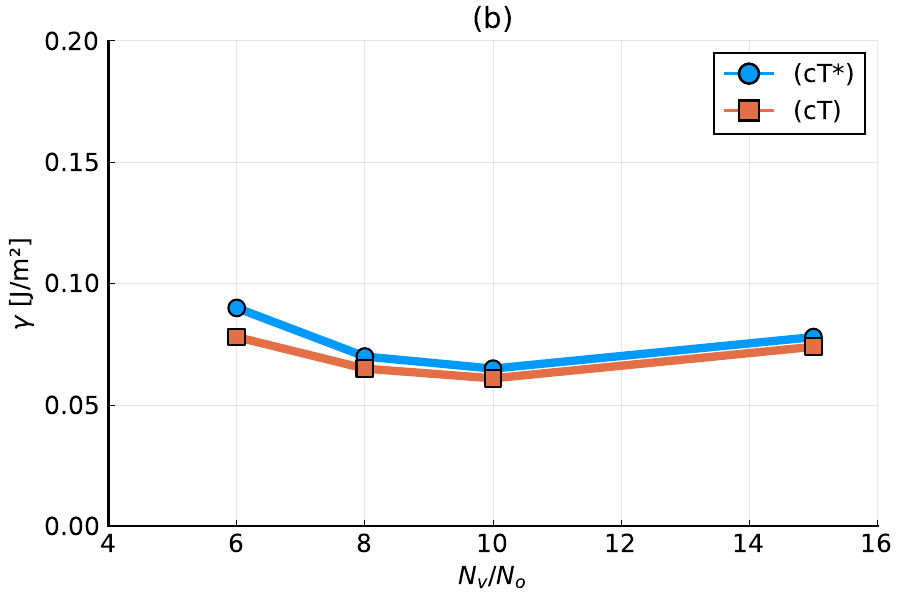}\\
    \includegraphics[width=0.49\linewidth]{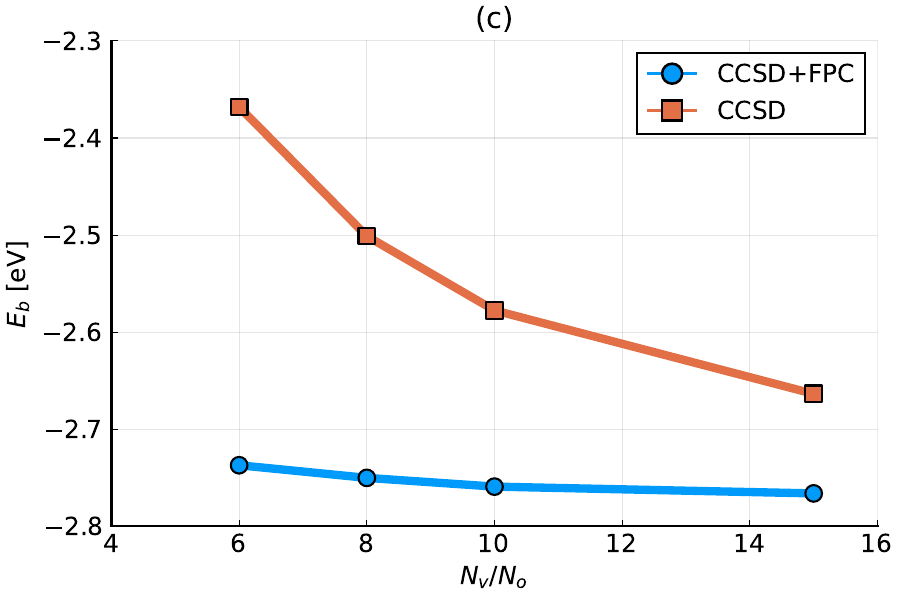}
    \includegraphics[width=0.49\linewidth]{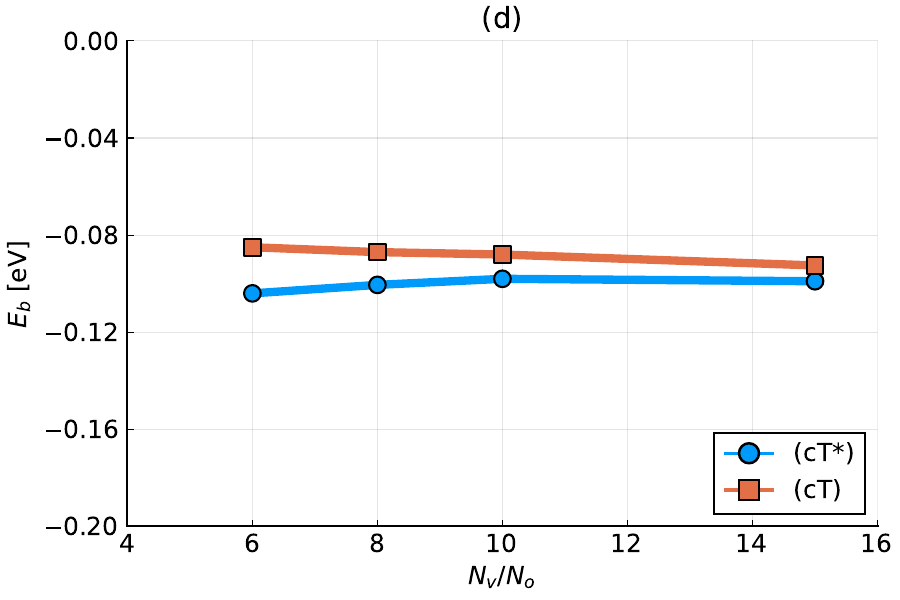}
    \caption{Basis-set convergence of the fit parameters $\gamma$ and $E_b$ of the fit $E(N_\text{layer}) = 2A\gamma + N_\text{layer}E_b$ for aluminum. A single twist angel was used. Only $N_\text{layer}=4,6$ was considered in order to directly capture the basis-set effect without averaging effects from multiple layers.}
    \label{fig:surfacecbs}
\end{figure}

Several procedures to calculate surface energies of periodic slabs were proposed in the literature.
In this work, we employ the fitting technique \cite{Mihm2023}, where the surface energy $\gamma$ is found by fitting the ground-state energies $E(N_\text{layer})$ against the number layers $N_\text{layer}$ through
\begin{equation}
E(N_\text{layer}) = 2\cdot A\cdot \gamma + N_\text{layer} \cdot E_b \;.
\end{equation}
Here, $A$ is the surface area of the slab and the other fitting parameter, $E_b$, can be interpreted as an estimate for the bulk energy.

For aluminum we considered $N_\text{layer} = 2, 4, 6, 8$.
As one aluminum atom provides an odd number of electrons, i.e. 3, in our setting, only even numbers of layers are chosen to ensure an even number of electrons in the $3\times 3$  slab.
For platinum  we considered $N_\text{layer} = 2, 3, 4, 5$, as one platinum atom provides an even number of electrons, i.e. 10.
Note that it was already shown in Ref. \cite{Miller2009} that the Pt(111) surface energy converges fast with respect to the number of layers using the fitting technique.

The \gls{hf} contribution for both Al and Pt was calculated using a $n\times n \times 1$ sampling of the \gls{bz} with $n=12$ and additionally averaging over stochastic twist angles, carefully checking for convergence with $n$ and the number of twist angles.

\subsection{Complete basis set limit of the surface energy}

The basis-set dependence of the surface energy of aluminum is shown in Fig. \ref{fig:surfacecbs}.
The highly effective focal point correction~\cite{Irmler2019} allows us to use $N_v/N_o = 8$ to achieve CCSD surface energies with a remaining uncertainty of less than $0.05\,\text{J/m}^2$.
Note that even $N_v/N_o = 15$ does not provide the same accuracy without the basis-set correction.
For the triples contributions, (cT), we find a much weaker basis-set dependence.
The choice of $N_v/N_o = 8$ introduces an even smaller error, especially when the (cT*) correction is used, as explained in Eq. (\ref{eq:star}).

\section{Finite-size correction scheme applied to bulk and surface}

\begin{figure}
    \centering
    \includegraphics[width=0.49\linewidth]{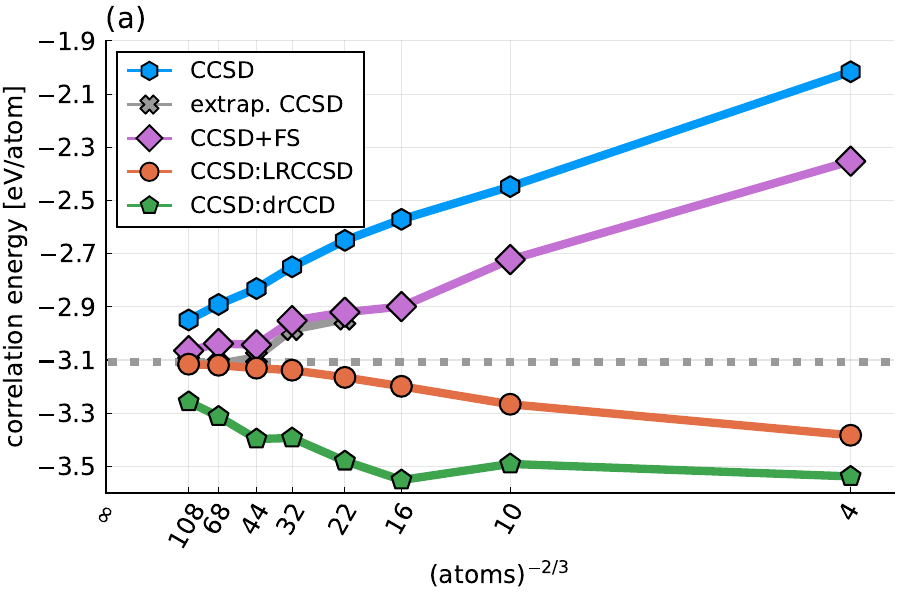}
    \includegraphics[width=0.49\linewidth]{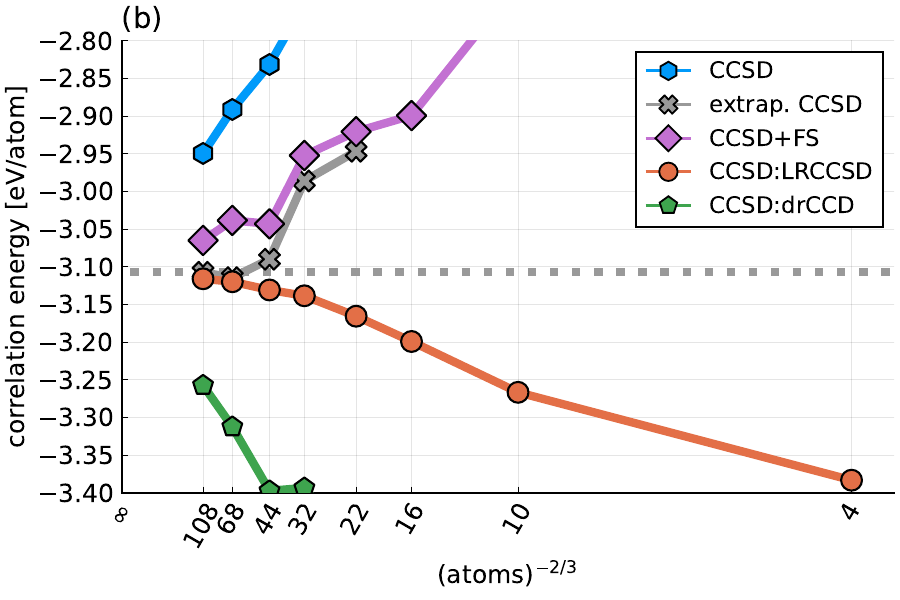}\\
    \includegraphics[width=0.49\linewidth]{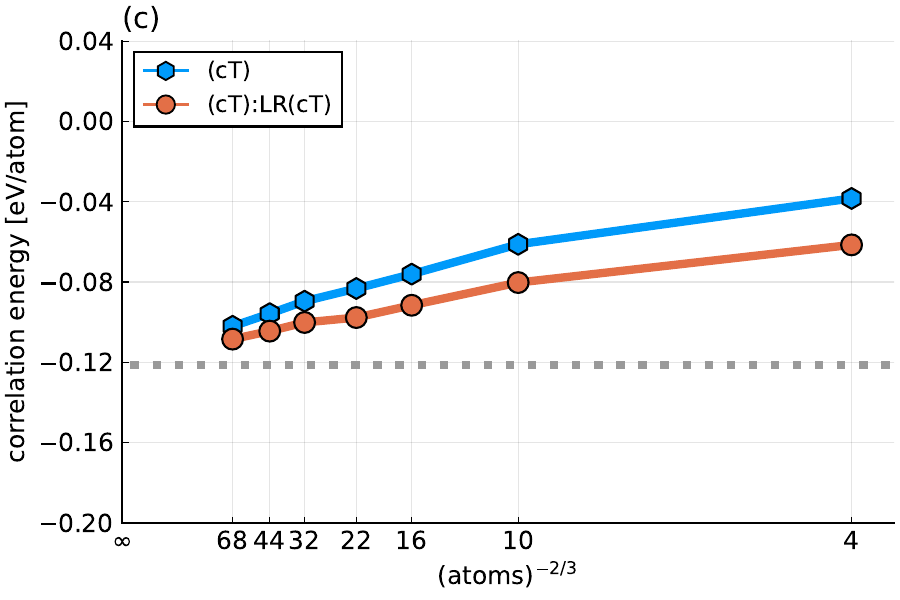}
    \caption{Approaching the total correlation energy per atom of metallic aluminum at the level of (a)\&(b) \gls{ccsd} and (c) (cT) using different finite-size corrections schemes. The correction schemes are introduced in the main text. The finite-size of the periodic model is indicated by the number of atoms on the horizontal axes. The extrapolation is performed using the largest four data points and is denoted as ``extrap. CCSD''. The best extrapolated CCSD / (cT) energy is shown as a dashed line. }
    \label{fig:EAl}
\end{figure}

\begin{figure}
    \centering
    \includegraphics[width=0.6\linewidth]{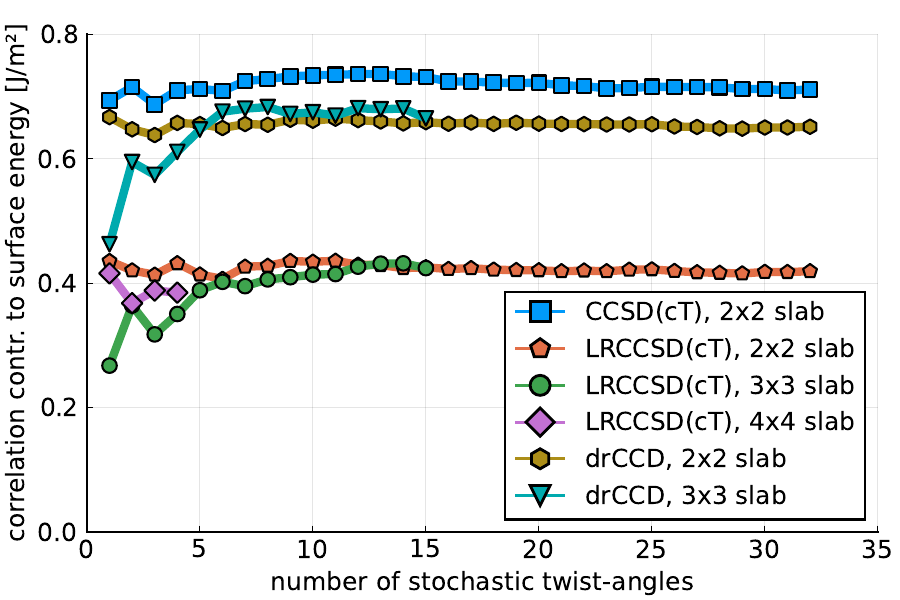}
    \caption{Dependence of the correlation contribution to the surface energy of Al(111) with respect to the number of stochastic shifts for the twist-averaging. Only for consistency we note that the CCSD(cT) as well as LRCCSD(cT) results are basis-set corrected while the drCCD results are not.}
    \label{fig:twistconverge}
\end{figure}

As defined in the main manuscript, the finite-size correction scheme shifts the task to reach the \gls{tdl} to the long-range coupled cluster (LRCC) part as
\begin{equation}
E^\text{TDL}_\text{CC} \approx  
E^\text{TDL}_\text{CC:LRCC} =
E^\text{finite}_\text{CC}
-E^\text{finite}_\text{LRCC}
+ E^\text{TDL}_\text{LRCC} \;.
\end{equation}
In this work we estimate $E^\text{TDL}_\text{LRCC}$ via cell size extrapolations based on cells with 32, 44, 68 and 108 atoms.
For the bulk an extrapolation low of $N^{-2/3}$ is assumed, where $N$ is the number of atoms per cell.
In Fig. \ref{fig:EAl} we show the convergence of the individual contributions to the correlation energy of bulk aluminum. 
Additionally, the convergence of the extrapolation technique is visualized.
%We note, that this is of minor relevance, as 

Considering the surface, Fig. \ref{fig:twistconverge} shows the correlation contribution from CCSD(cT) and LRCCSD(cT) to the surface energy using slabs growing in the $xy$ direction in dependence of the used stochastic shifts for the twist-averaging technique.
The stochastic noise of the twist-averaging leaves an uncertainty of about $0.1 \,\text{J/m²}$ for the $4\times 4$ slab.
For all other systems the uncertainty is smaller.
Within this error bar, the correlation contribution is already converged at the $2\times 2$ slabs.

To hedge this surprising result, we also considered the change of the \gls{drccd} surface energy from $2\times 2$ slabs to $3 \times 3$ slabs using the full Coulomb potential.
This is possible as \gls{drccd} is computationally less demanding than \gls{ccsd} or \gls{ccsdpct} calculations.
Clearly, the drCCD correlation contribution to the surface energy agree for both the $2\times 2$ slab and the $3\times 3$ slab within an error of less that $0.05 \,\text{J/m²}$.
In other words, the correlation contribution to the surface energy does indeed only have a very weak dependence on the $xy$ direction.
To keep the computationally cost as low as possible, we neglected a basis-set correction for the drCCD results of this test.

\bigskip

\begin{figure}
    \centering
    \includegraphics[width=0.6\linewidth]{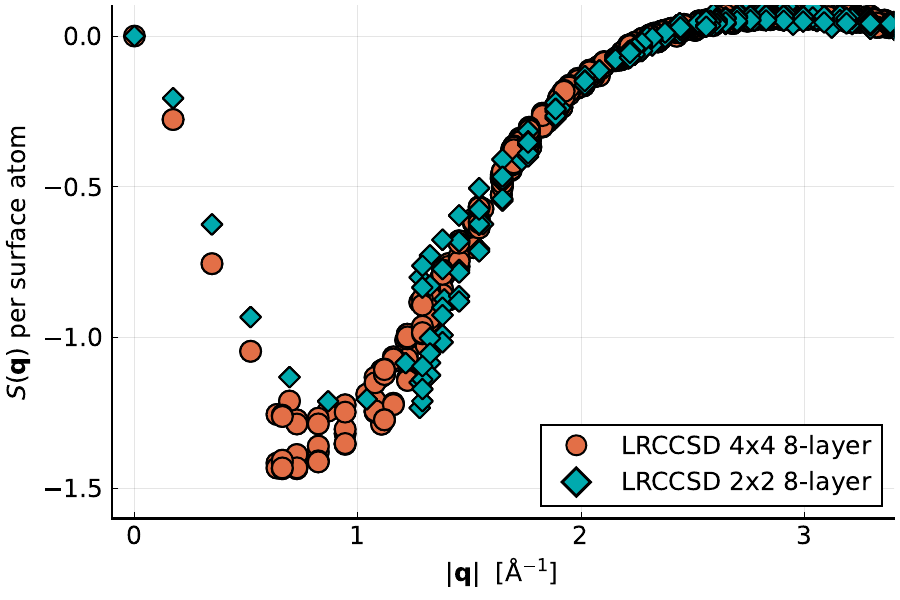}
    \caption{LRCCD structure factors of $n\times n$ aluminum surface slabs with 8 layers and $n=2,4$ using a single twist. Note that that the considered supercells are strongly anisotropic.}
    \label{fig:surfaceSFAl}
\end{figure}

Turning to the finite-size correction based on the structure factor interpolation~\cite{Liao2016}, we observe a massive underestimation of surface energies.
%We speculate that this is due to the strong anisotropy of the cell $2\times 2$ slabs.
As shown in Fig. \ref{fig:surfaceSFAl}, the $2 \times 2$ structure factor does not reach its characteristic minimum for vectors in the $xy$ direction. 
Hence, the interpolation introduces severe numerical errors, leading to the wrong CCSD(cT)+FS estimates for the surface energy, mentioned in the main manuscript

\section{Timings and computer architecture}

\begin{figure}
    \centering
    \includegraphics[width=0.6\linewidth]{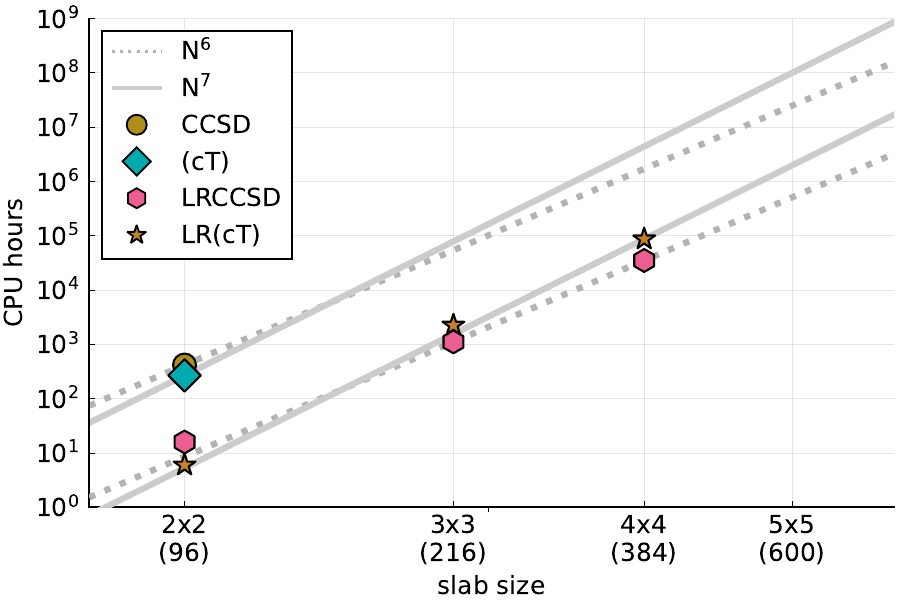}
    \caption{Log-log plot of the computation time for Aluminum slabs with 8 layers. Both CCSD timings are given per 20 iterations, providing a rough estimate for the total required. The number of electrons in the slab model is given in brackets.}
    \label{fig:timinig}
\end{figure}

Figure \ref{fig:timinig} shows the \gls{cc} computation time for 8-layer aluminum slabs.  
Up to 32 nodes with 128 cores each were used for the largest calculations. 
The nodes are equipped with AMD EPYC 7713 CPUs and 512 GB of main memory.
They are provided by the Vienna Scientific Cluster (VSC-5).

\bibliography{refs}% Produces the bibliography via BibTeX.

%% file: cmd.tex
\newcommand{\D}{\text d}

\newcommand{\EuE}{\text e}

\newacronym{hf}{HF}{Hartree-Fock}
\newacronym{cc}{CC}{coupled cluster}
\newacronym{ccsd}{CCSD}{coupled cluster with single and double particle–hole excitation operators}
\newacronym{ccsdpct}{CCSD(cT)}{CCSD(cT)}
\newacronym{drccd}{drCCD}{direct-ring coupled cluster doubles}
\newacronym{ccsdpt}{CCSD(T)}{coupled cluster with single, double and perturbative triple particle–hole excitation operators}
\newacronym{bz}{BZ}{Brillouin zone}
\newacronym{tdl}{TDL}{thermodynamic limit}
\newacronym{si}{SI}{supporting information}
\newacronym{cbs}{CBS}{complete basis set limit}
\newacronym{dft}{DFT}{density functional theory}
\newacronym{rpa}{RPA}{random phase approximation}
\newacronym{vasp}{VASP}{Vienna Ab-initio Simulation Package}
\newacronym{paw}{PAW}{projector augmented wave}